\documentclass[letterpaper,twoside,journal,final,twosided]{IEEEtran}
\usepackage[T1]{fontenc}
\usepackage[latin9]{inputenc}
\usepackage{float}
\usepackage{amsmath}
\usepackage{amssymb}
\usepackage{graphicx}
\usepackage[unicode=true,pdfusetitle,
 bookmarks=true,bookmarksnumbered=false,bookmarksopen=false,
 breaklinks=false,pdfborder={0 0 1},backref=false,colorlinks=false]
 {hyperref}
\usepackage{breakurl}

\makeatletter

\setkeys{Gin}{width=\linewidth}
\usepackage{cite}

\long\def\@makecaption#1#2{\ifx\@captype\@IEEEtablestring%
\footnotesize\begin{center}{\normalfont\footnotesize #1}\\
{\normalfont\footnotesize\scshape #2}\end{center}%
\@IEEEtablecaptionsepspace
\else
\@IEEEfigurecaptionsepspace
\setbox\@tempboxa\hbox{\normalfont\footnotesize {#1.}~~ #2}%
\ifdim \wd\@tempboxa >\hsize%
\setbox\@tempboxa\hbox{\normalfont\footnotesize {#1.}~~ }%
\parbox[t]{\hsize}{\normalfont\footnotesize \noindent\unhbox\@tempboxa#2}%
\else
\hbox to\hsize{\normalfont\footnotesize\hfil\box\@tempboxa\hfil}\fi\fi}
\ifCLASSOPTIONcompsoc
  \usepackage[caption=false,font=normalsize,labelfont=sf,textfont=sf]{subfig}
\else
  \usepackage[caption=false,font=footnotesize]{subfig}
\fi

\DeclareMathOperator*{\argmax}{arg\,max}

\usepackage{dblfloatfix}

\@ifundefined{showcaptionsetup}{}{%
 \PassOptionsToPackage{caption=false}{subfig}}
\usepackage{subfig}
\makeatother

\begin{document}

\title{Inter-numerology Interference Management with~Adaptive Guards: A
Cross-layer Approach}

\author{Ali Fatih Demir, \emph{Student Member, IEEE}, H{\"u}seyin Arslan, \emph{Fellow,
IEEE}\vspace{-0mm}
\thanks{Manuscript received December 13, 2019; accepted January 07,
2020.}\thanks{Ali Fatih Demir is with the Department of Electrical
Engineering, University of South Florida, Tampa, FL 33620 USA (e-mail:
afdemir@mail.usf.edu).}\thanks{H{\"u}seyin Arslan is with the Department
of Electrical Engineering, University of South Florida, Tampa, FL
33620 USA, and also with the Department of Electrical and Electronics
Engineering, Istanbul Medipol University, Istanbul 34810, Turkey (e-mail:
arslan@usf.edu).}}
\maketitle
\begin{abstract}
The next-generation communication technologies are evolving towards
increased flexibility in various aspects. Although orthogonal frequency
division multiplexing (OFDM) remains as the waveform of the upcoming
fifth-generation (5G) standard, the new radio provides flexibility
in waveform parametrization (a.k.a. numerology) to address diverse
requirements. However, managing the peaceful coexistence of mixed
numerologies is challenging due to inter-numerology interference (INI).
This paper proposes the utilization of adaptive guards in both time
and frequency domains as a solution along with a multi-window operation
in the physical (PHY) layer. The adaptive windowing operation needs
a guard duration to reduce the unwanted emissions, and a guard band
is required to handle the INI level on the adjacent band. The guards
in both domains are jointly optimized with respect to the subcarrier
spacing, use case (i.e., service requirement), and power offset between
the numerologies. Also, the multi-window approach provides managing
each side of the spectrum independently in case of an asymmetric interference
scenario. Since the allowed interference level depends on the numerologies
operating in the adjacent bands, the potential of adaptive guards
is further increased and exploited with a medium access control (MAC)
layer scheduling technique. The proposed INI-based scheduling algorithm
decreases the need for guards by allocating the numerologies to the
available bands, considering their subcarrier spacing, power level,
and SIR requirements. Therefore, INI management is performed with
a cross-layer (PHY and MAC) approach in this study. The results show
that the precise design that accommodates such flexibility reduces
the guards significantly and improves the spectral efficiency of mixed
numerology systems.
\end{abstract}


\markboth{IEEE Access}{Demir \MakeLowercase{and} Arslan: Inter-numerology Interference Management with Adaptive Guards: A Cross-layer Approach}

\section{Introduction \label{sec:I}}

\begin{figure*}
\centering\includegraphics[width=2\columnwidth]{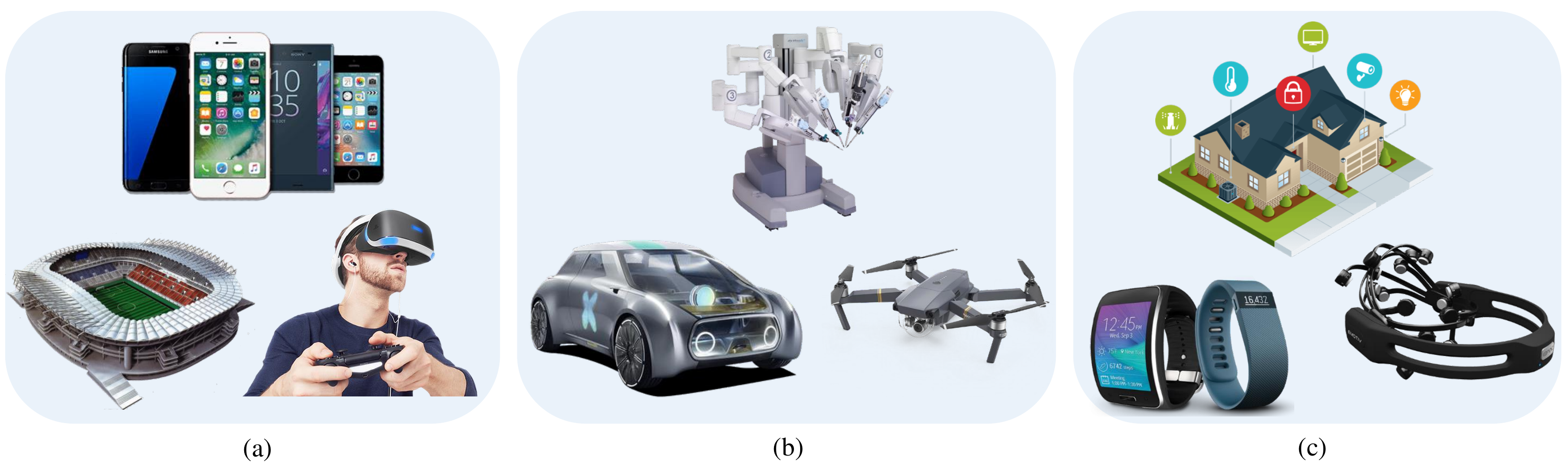}\caption{5G use cases: (a) Enhanced mobile broadband, (b) ultra-reliable low-latency
communications, and (c) massive machine type communications. \label{fig:5G-Use-Cases}}
\end{figure*}

\IEEEPARstart{T}{he} next generation wireless communication technologies
are envisioned to support a diverse range of services under the same
network. As a recent example, the International Telecommunications
Union (ITU) has defined the main use cases that are going to be supported
in the fifth generation (5G) mobile network as enhanced mobile broadband
(eMBB), massive machine type communications (mMTC), and ultra-reliable
low-latency communications (URLLC) \cite{ITU-R} as shown in Fig.
\ref{fig:5G-Use-Cases}. The applications which demand high data rate
and better spectral efficiency fall into the eMBB category, whereas
the ones which require ultra-high connection density and low power
consumption falls into the mMTC category. Moreover, the mission-critical
applications, where errors and retransmissions are less tolerable,
are categorized in URLLC. A flexible air interface is needed to meet
these demanding service requirements under various channel conditions
and system scenarios. Hence, the waveform, which is the main component
of any air interface, must be designed precisely to facilitate such
flexibility.

Orthogonal frequency-division multiplexing (OFDM) is the most popular
waveform that is currently being used in various standards such as
4G LTE and the IEEE 802.11 family \cite{hwang2009}. It provides several
tempting features such as efficient hardware implementation, low-complexity
equalization, and easy multiple-input-multiple-output (MIMO) integration.
On the other hand, OFDM seriously suffers from its high out-of-band
emissions (OOBE), peak-to-average power ratio (PAPR), and strict synchronization
requirements. In addition, 4G LTE adopts a uniform OFDM parameter
configuration in pursuit of orthogonality and cannot serve different
needs efficiently. Numerous waveforms have been proposed \cite{demir2018,zhang2016,3gppQualCand,sahin2016,berardinelli2016}
considering all these disadvantages, but none of them can address
all the requirements of the upcoming 5G standard \cite{3gppHuaweiReq}.
Therefore, OFDM remains as the waveform of the new radio \cite{kim2019,parkvall2017,lien2017},
and a flexible waveform parametrization, which is also known as numerology
\cite{ankarali2017,zaidi2016}, is introduced to embrace diverse requirements.

The channel conditions, use cases, and system scenarios are the most
critical considerations for the numerology design. For example, the
subcarrier spacing of OFDM should be kept large to handle the Doppler
spread in a highly mobile environment. On the other hand, a smaller
subcarrier spacing provides a longer symbol duration and decreases
the relative redundancy that is allocated for time dispersion. An
efficient numerology design ensures better utilization of spectral
resources and numerology multiplexing will be one of the core technologies
in the new radio \cite{3gppHuaweiNum}. However, managing the coexistence
of multiple numerologies in the same network is challenging. Although
OFDM numerologies are orthogonal in the time domain, any mismatch
in parametrization, such as subcarrier spacing, leads to inter-numerology
interference (INI) in the frequency domain \cite{ankarali2017,kihero2019}.
Despite the fact that it is a new interference type, which will be
an issue for 5G, extensive research led to in-depth INI analyses and
various INI management techniques \cite{kihero2019,zhang2018,pekoz2019,yang2019,levanen2019,yli-kaakinen2017,zhang2017,lim2018,ankarali2017}.
For instance, windowing \cite{zhang2018,pekoz2019} and filtering
\cite{yang2019,yli-kaakinen2017} operations are performed at both
transmitter and receiver side along with the guard band allocation
to mitigate the unwanted emissions from non-orthogonal numerologies.
In addition, it is demonstrated that mixed transparent waveform processing
can be applied to optimize the complexity-performance trade-off at
transmitter and receiver separately \cite{levanen2019}. Precoding
techniques at the transmitter \cite{zhang2017,huang2018} and interference
cancellation algorithms at the receiver \cite{zhang2018} are also
considered for multi-numerology coexistence. Last but not least, waveform
multiplexing \cite{ankarali2017,lim2018} is suggested for mixed numerology
management as well. Among these physical (PHY) layer techniques, the
windowing operation has a relatively less complexity, which is almost
at the same level compared to CP-OFDM \cite{levanen2019}. Also, the
windowing approach preserves the essential structure of the OFDM transceivers
and provides backward compatibility for the current OFDM-based systems.
The windowing operation requires an extra period, which extends the
guard duration between the consecutive OFDM symbols. Also, additional
guard bands are still required between adjacent channels to deal with
the INI. In other words, better interference mitigation is realized
with the cost of reduced spectral efficiency. Accordingly, the future
communication systems have to optimize the guards in both time and
frequency domains to improve the spectral efficiency.

This paper proposes the utilization of adaptive guards along with
a multi-window operation in the PHY layer to manage the INI, which
is an issue in the mixed numerology systems. The guard band and the
window parameters that handle the guard duration are jointly optimized
regarding the subcarrier spacing, use case, and power offset between
the numerologies. Also, the multi-window technique provides managing
each side of the spectrum independently in case of an asymmetric interference
scenario. Since the allowed interference level depends on the numerologies
operating in the adjacent bands, the potential of adaptive guards
is further increased and exploited with a medium access control (MAC)
layer scheduling technique. The proposed INI-based scheduling algorithm
decreases the need for guards by allocating the numerologies to the
available bands, considering the subcarrier spacing, power level,
and SIR requirements. Therefore, INI management is performed with
a cross-layer (PHY and MAC) approach in this study. The preliminary
results without a mixed-numerology guard optimization, a multi-window
operation, or an elaborate INI-based scheduling algorithm and its
evaluation were presented in \cite{demir2017}. Recently, a U.S. patent
\cite{demir2019patent} is issued for the proposed technique as well.
The main contributions of this paper are listed as follows:
\begin{itemize}
\item The key parameters for guard allocation are identified considering
a mixed numerology system. 
\item The guards in both time and frequency domains are jointly optimized
with respect to the subcarrier spacing, use case, and power offset
between the numerologies. 
\item An interference based scheduling algorithm is proposed to decrease
the need for guards.
\end{itemize}
The remaining part of this paper is structured as follows. Section
\ref{sec:II} is dedicated to the system model, and it describes the
guard design methodology in detail. Section \ref{sec:III} presents
the guard optimization procedure considering the key parameters of
the mixed numerology system. Section \ref{sec:IV} introduces the
INI-based scheduling algorithm along with the utilization of adaptive
guards. Finally, Section \ref{sec:V} summarizes the contributions
and concludes the paper. 

\section{System Model \label{sec:II}}

\begin{figure}
\centering\includegraphics[width=1\columnwidth]{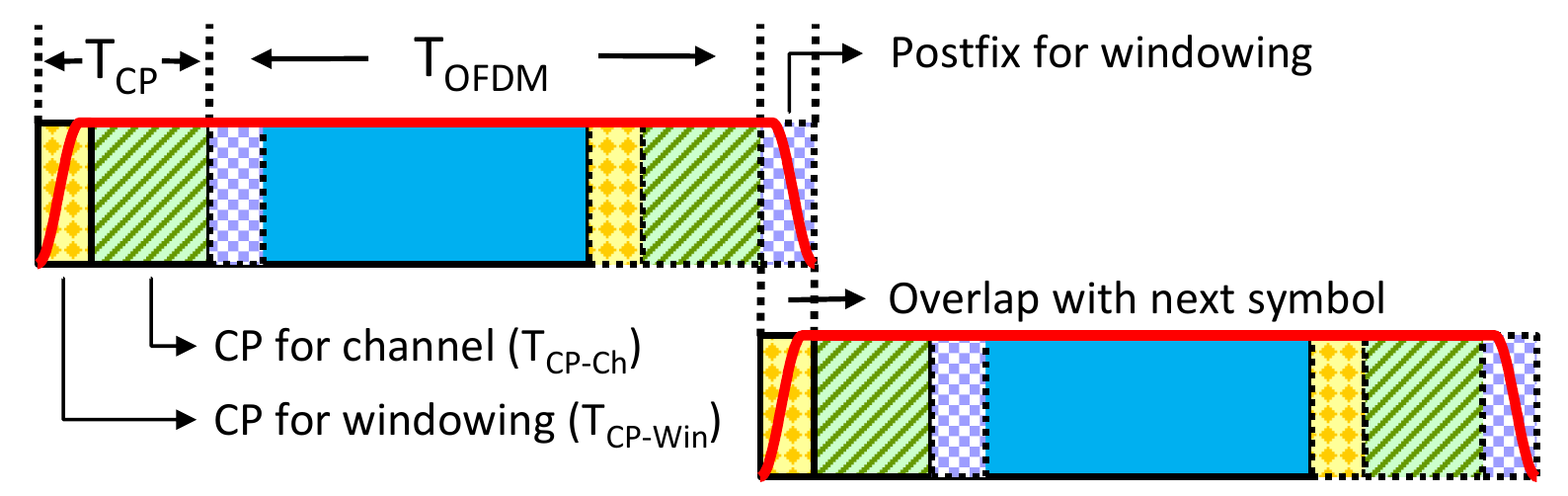}
\centering{}\caption{Transmitter windowing operation and the guard duration allocation.
\label{fig:GuardDuration}}
\end{figure}

Consider the uplink of a multiuser OFDM system, where asynchronous
numerologies with different subcarrier spacing, power level, and use
case (i.e., service requirements) operate in the same network. Each
numerology can serve multiple synchronous user equipments (UEs) and
is assigned to a different bandwidth part (BWP) \cite{parkvall2017}.
A transmitter windowing operation is performed to improve the spectral
localization of numerologies and manage interference level on the
adjacent BWPs. The guard duration that is allocated for the time-dispersive
channel (i.e., $T_{CP-Ch}$) is fixed, and it is adequate to deal
with the inter-symbol interference (ISI). Also, an extra guard duration
is needed for windowing operation. Various windowing functions have
been compared thoroughly \cite{farhang2011} with different trade-offs
between the main lobe width and the side lobe suppression. The optimal
windowing function is outside the scope of this paper, and a raised-cosine
(RC) window is utilized due to its low computational complexity and
widespread use in the literature \cite{weiss2004,bala2013,sahin2014}.
The RC window function \cite{weiss2004} is formulated by the following
equation:

\begin{equation}
\hspace{-8.5pt}g[n]=\begin{cases}
\frac{1}{2}+\frac{1}{2}\cos\left(\pi+\frac{\pi n}{\alpha N_{\textrm{T}}}\right) & 0\leq n\leq\alpha N_{\textrm{T}}\\
1 & \alpha N_{\textrm{T}}\leq n\leq N_{\textrm{T}}\\
\frac{1}{2}+\frac{1}{2}\cos\left(\pi-\frac{\pi n}{\alpha N_{\textrm{T}}}\right) & N_{\textrm{T}}\leq n\leq\left(\alpha+1\right)N_{\textrm{T}}
\end{cases}\label{eq:RC_Window}
\end{equation}
where $\alpha$ is the roll-off factor ($0\leq\alpha\leq1$) and $N_{\textrm{T}}$
denotes the symbol length. The roll-off factor ($\alpha$) handles
the taper duration of the RC window function. As $\alpha$ increases,
the INI decreases with the cost of increased redundancy. The transmitter
windowing operation is shown in Fig. \ref{fig:GuardDuration}. Initially,
the guard duration is increased with an additional cyclic prefix (CP)
and a newly added cyclic suffix (CS). Afterward, the window function
is applied to the extended symbol. The transition parts (i.e., ramp-ups
and ramp-downs) of adjacent symbols are overlapped to reduce the time-domain
overhead emerging from the windowing operation.

\begin{figure}[b]
\centering\includegraphics[width=0.68\columnwidth]{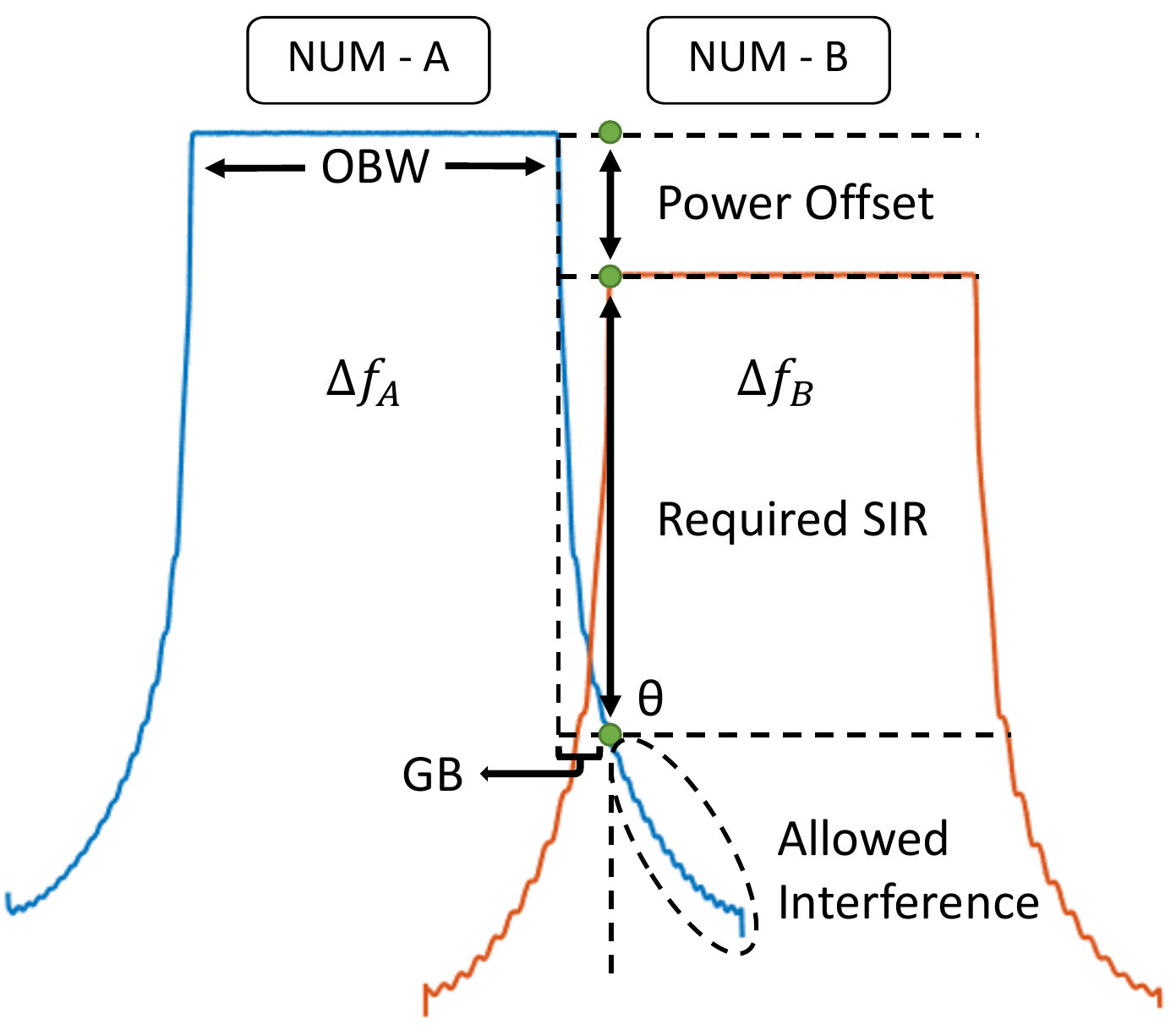}
\centering{}\caption{Guard band allocation between two numerologies considering the allowed
interference level ($\theta$) in the adjacent band. \label{fig:GuardBand}}
\end{figure}

\begin{figure}[t]
\centering\includegraphics[width=1\columnwidth]{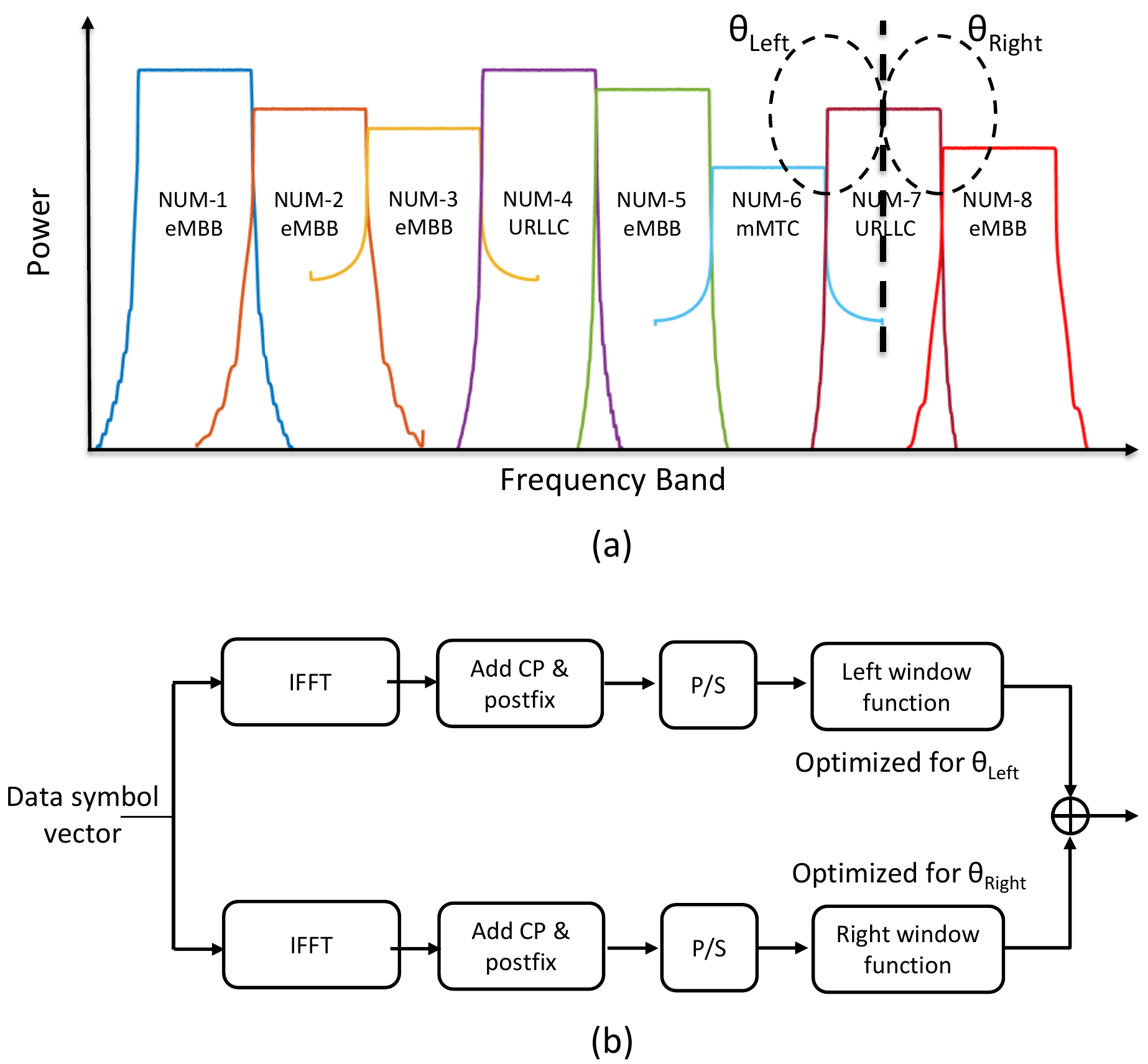}
\centering{}\caption{(a) Asymmetric interference scenario in a mixed numerology network;
(b) Block diagram of the multi-window operation. \label{fig:Multi-window}}
\end{figure}

Usually, the windowing operation is not enough to manage the inter-numerology
interference (INI), and non-negligible guard bands are still required.
However, the total amount of guard band (GB) or the length of guard
duration (GD) which is needed for windowing operation depends on the
subcarrier spacing of the interference source, the required signal
to interference ratio (SIR) level of the numerology in its adjacent
bands, and the power offset (PO) between them. The adaptive guard
concept is represented with two numerologies in Fig. \ref{fig:GuardBand}
and can be generalized to multiple numerologies by considering one
pair at a time. The threshold for allowed interference level on the
adjacent band is represented with $\theta$, and it is expressed as
follows: 

\begin{equation}
\theta_{\Delta f,i}=P_{i}-P_{j}+S_{j}\label{eq:Int_Threshold}
\end{equation}
where $P_{i}$ represents the in-band power of the interference source,
$S_{j}$ denotes the required SIR in the adjacent band to achieve
a given target bit error rate considering device complexity for processing,
and $\Delta f$ indicates the subcarrier spacing of the interference
source. It should be noted that the different use case requirements
and device capabilities are reflected in the required SIR parameter,
power level, and subcarrier spacing. Also, $\theta_{\Delta f}$ implies
an adaptive brick-wall type spectral mask for a simple evaluation
in this study. However, it is possible to extend it to more complicated
mask structures along with the adjacent channel leakage ratio (ACLR)
threshold \cite{3gppTS}. The guards in both time and frequency domains
are utilized regarding $\theta_{\Delta f}$ to achieve the desired
SIR level of the numerology on the adjacent band. Throughout the numerical
evaluations in this study, GD (i.e., $T_{CP-Win}$) and GB are adaptive,
and these guards are optimized in Section \ref{sec:III}. Also, a
multi-window operation \cite{sahin2011a}\cite{3gppSamsungEdge} can
be performed in case of an asymmetric interference scenario to manage
each side of the spectrum independently considering $\theta_{Left}$
and $\theta_{Right}$ as shown in Fig. \ref{fig:Multi-window}. The
total CP length must be kept the same for synchronicity when a multi-window
operation is performed, and the extra CP duration is reserved to solve
possible time-domain issues. The remaining parameters of the windowed-OFDM
(W-OFDM) system are listed in Table \ref{tab:Sim-Par}.

\begin{table}[H]
\caption{Simulation Parameters \label{tab:Sim-Par}}

\centering\includegraphics[width=0.9\columnwidth]{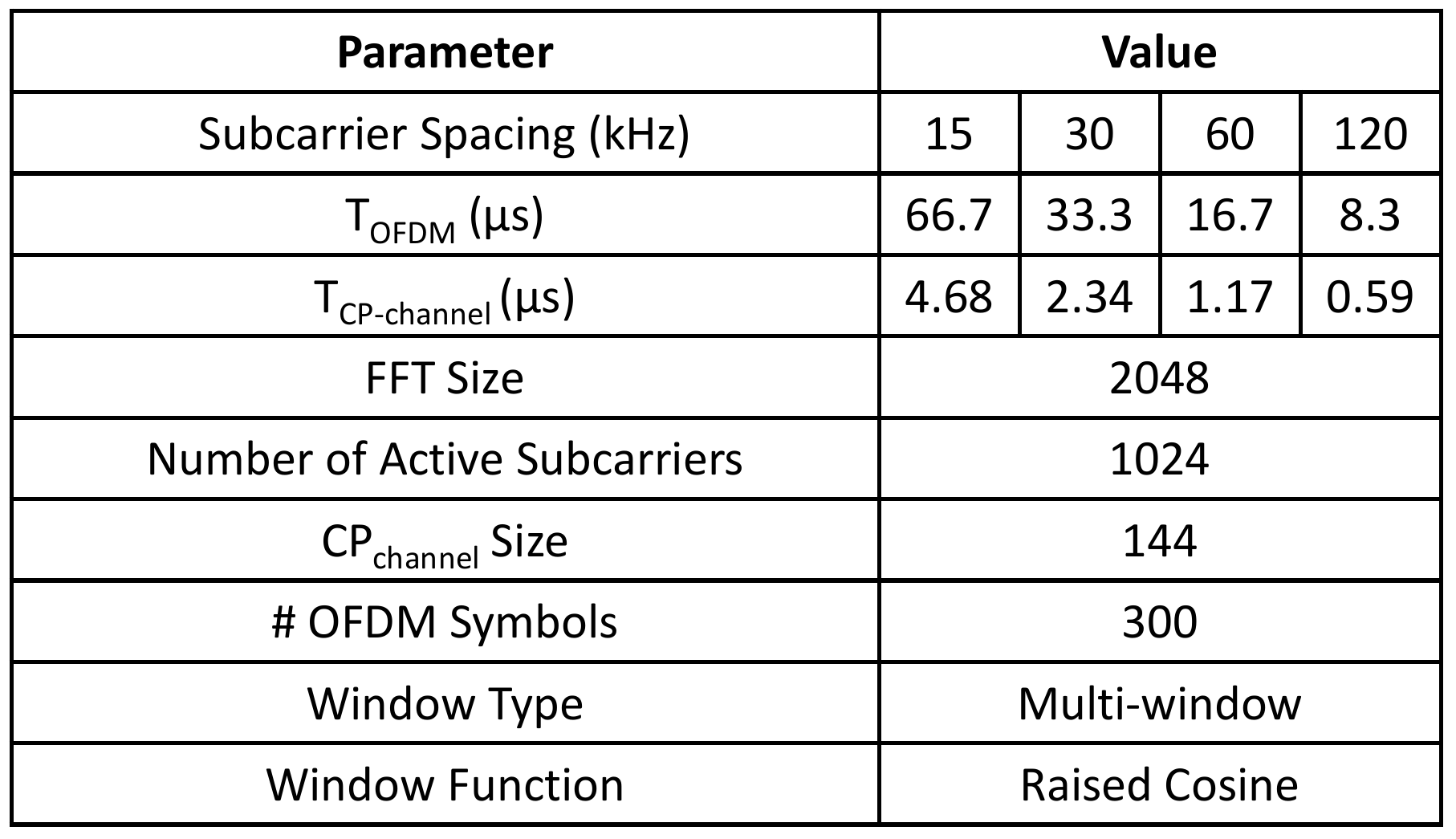}
\end{table}

\begin{figure}[t]
\centering\includegraphics[width=1\columnwidth]{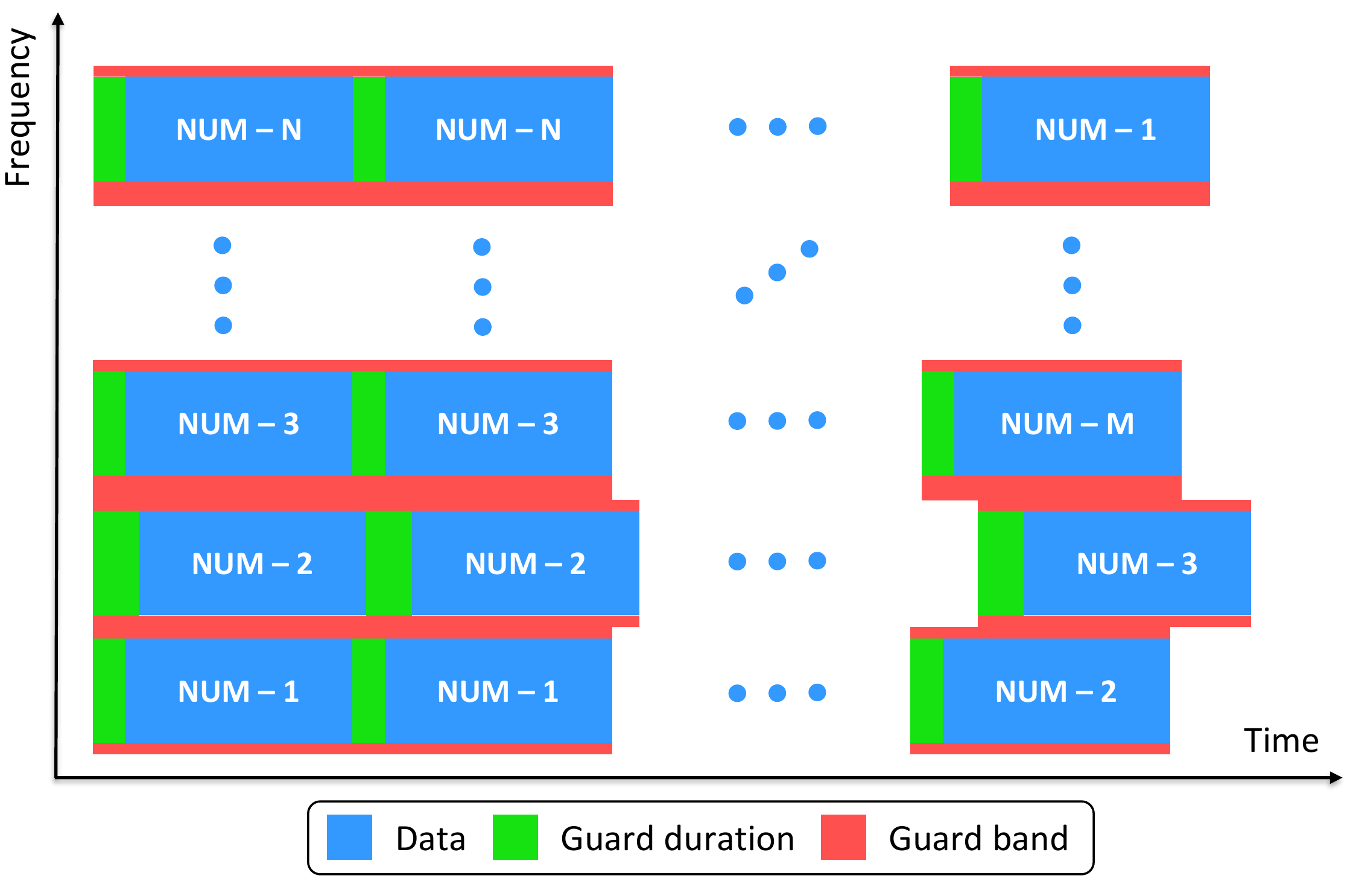}
\centering{}\caption{Frequency domain multiplexed numerologies. \label{fig:FDM-Numerologies}}
\end{figure}

The potential of adaptive guards is increased further, along with
the utilization of INI-based scheduling algorithm. Consider frequency
domain multiplexed $M$ asynchronous numerologies as shown in Fig.
\ref{fig:FDM-Numerologies}. Different channel conditions, use cases,
and system scenarios result in a change in subcarrier spacing, power
level, and SIR requirement of the numerologies as mentioned in Section
\ref{sec:I}. The optimal numerology assignment is beyond the scope
of this study, and the reader is referred to \cite{yazar2018} for
more details on this topic. In this article, the spectral efficiency
is optimized while ensuring the required SIR levels for a given numerology
set. The power level and SIR requirement of each numerology are generated
randomly in such a way that $\theta$ changes from $0$ dB to $60$
dB. Also, $\Delta f$ gets discrete values of $\left\{ 15,30\right\} $
kHz and $\left\{ 60,120\right\} $ kHz with a uniform probability
distribution in the frequency range-1 (FR1, a.k.a. sub-6 GHz bands)
and frequency range-2 (FR2, a.k.a. millimeter-wave bands) \cite{kim2019},
respectively. Assuming that the base station obtains all these necessary
information perfectly and there are $M$ available subbands, it allocates
the numerologies to the available subbands out of $M!$ possible arrangements
intelligently considering the INI.

\section{Optimization of The Adaptive Guards\label{sec:III}}

\begin{figure}
\subfloat[]{\centering\includegraphics[width=1\columnwidth]{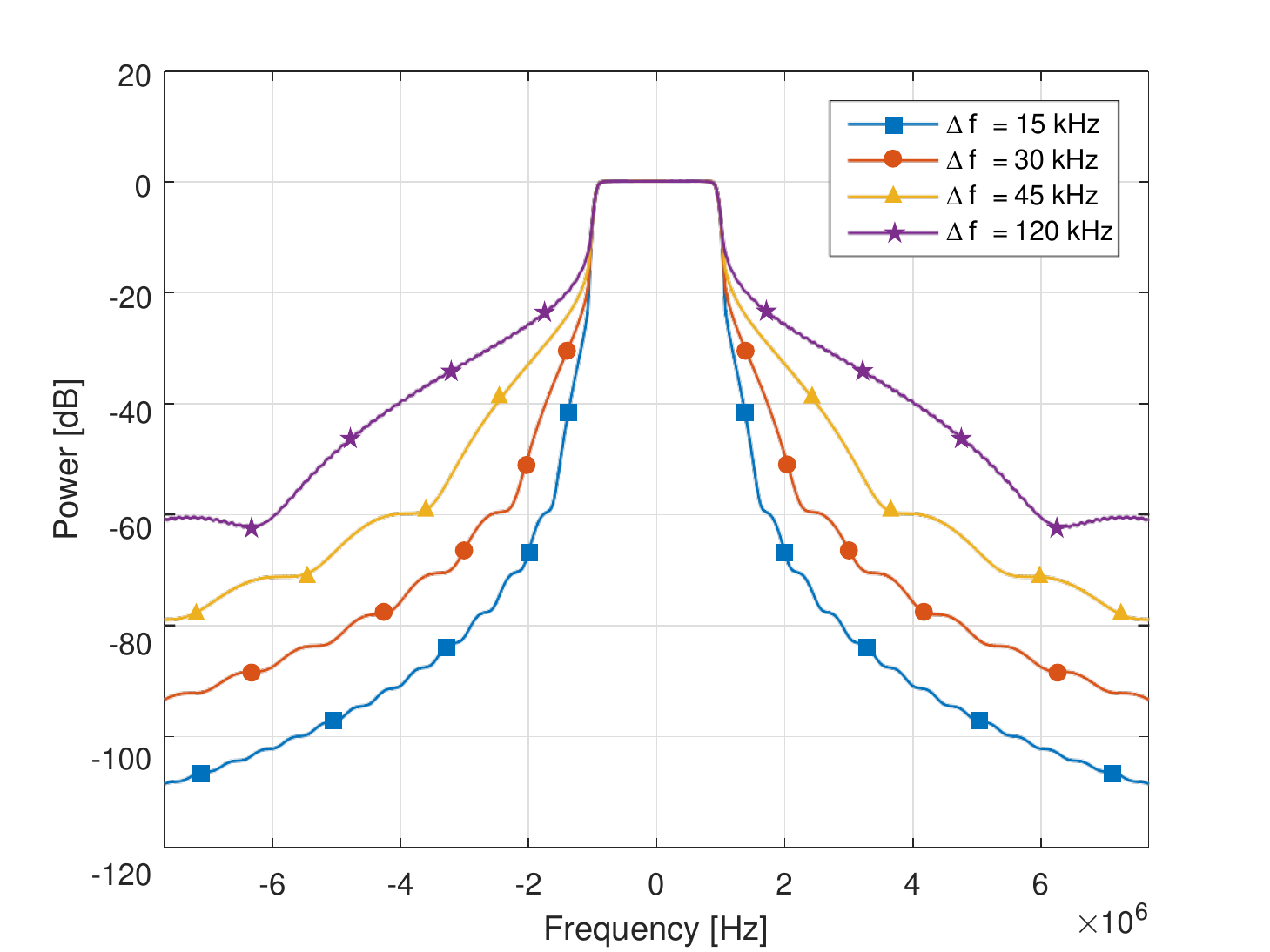}

}

\subfloat[]{\centering\includegraphics[width=1\columnwidth]{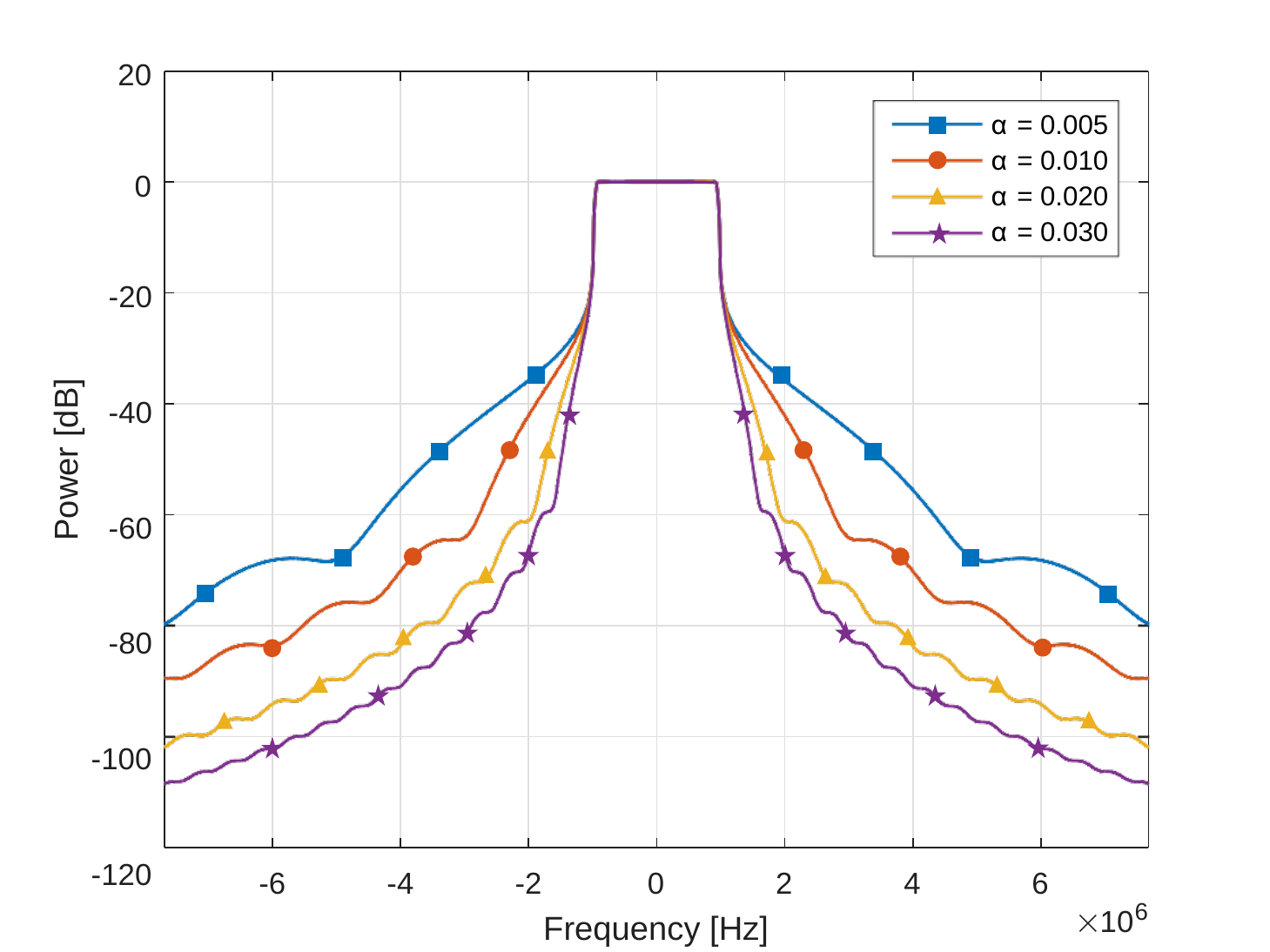}

}

\caption{PSD of W-OFDM: (a) The effect of $\Delta f$ ($\alpha$ is fixed to
$0.03$); (b)~The effect of $\alpha$ ($\Delta f$ is fixed to $15$
kHz). \label{fig:PSD-Parameters}}
\end{figure}
Assuming that the data at each subcarrier are statistically independent
and mutually orthogonal, the power spectral density (PSD) of an OFDM
signal is obtained by summing the power spectra of individual subcarriers,
and it is expressed by the following equation \cite{waterschoot2010,talbot2008,liu2004}:

\begin{equation}
P_{f}(x)=\frac{\sigma_{d}^{2}}{T}\sum_{k}\left|G\left[\left(f-k\Delta f\right)T\right]\right|^{2}\label{eq:PSD_of_OFDM}
\end{equation}
where $\sigma_{d}^{2}$ represents the variance of the data symbols,
$T$ denotes the symbol duration, $k$ stands for the subcarrier index,
$\Delta f$ shows the subcarrier spacing, and $G$ is the frequency
domain representation of pulse shaping window. An OFDM signal is well
localized in the time domain with a rectangular pulse shape, which
is equivalent to a sinc shape in the frequency domain. The sidelobes
of the sincs result in a serious INI issue, and they should be reduced
to prevent interference. Particularly, the frequency domain localization
is crucial for asynchronous transmissions across adjacent subbands
and peaceful coexistence with other numerologies in the network. The
sidelobes of RC function is controlled with the parameter $\alpha$
as shown in the following relationship \cite{broadcom2013}:
\begin{figure}[t]
\centering\includegraphics[width=1\columnwidth]{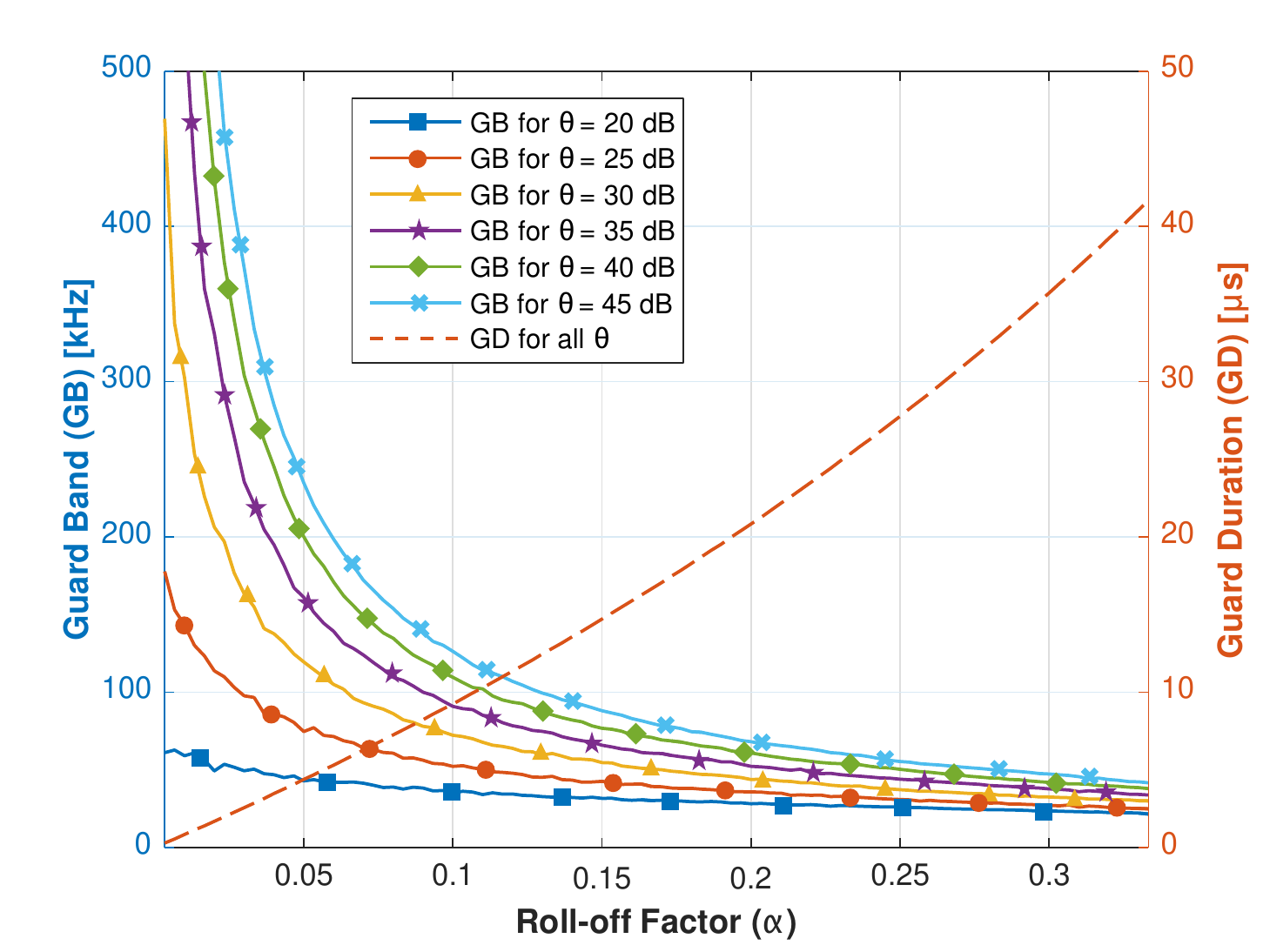}
\centering{}\caption{Required GB and GD pairs to achieve selected $\theta$ levels for
a W-OFDM signal with $\Delta f=15\:kHz$. \label{fig:GBvsGD}}
\end{figure}

\begin{equation}
G=\frac{\sin\left(\pi fT\right)}{\pi fT}\:\frac{\cos\left(\pi\alpha fT\right)}{1-\left(2\alpha fT\right)^{2}}\:\:\:\:\:\:\:\:\:\:\:\:0\leq\alpha\leq1\label{eq:PSD_of_RC}
\end{equation}
Equation \ref{eq:PSD_of_OFDM} and equation \ref{eq:PSD_of_RC} show
that the parameters $T$ (i.e., $\Delta f=1/T$) and $\alpha$ have
an important effect on the PSD of W-OFDM. Figure \ref{fig:PSD-Parameters}
illustrates the effect of these parameters on the PSD separately.
It should be noted that a significant contribution to unwanted emissions
in the passband comes from RF front-end impairments as well, including
power amplifier nonlinearities. However, these impairments heavily
depend on many implementation-dependent factors, such as the application
type, operational frequency, bandwidth of the signal, and complexity
of the device, and are not considered in this study.

\begin{figure}[t]
\centering\includegraphics[width=1\columnwidth]{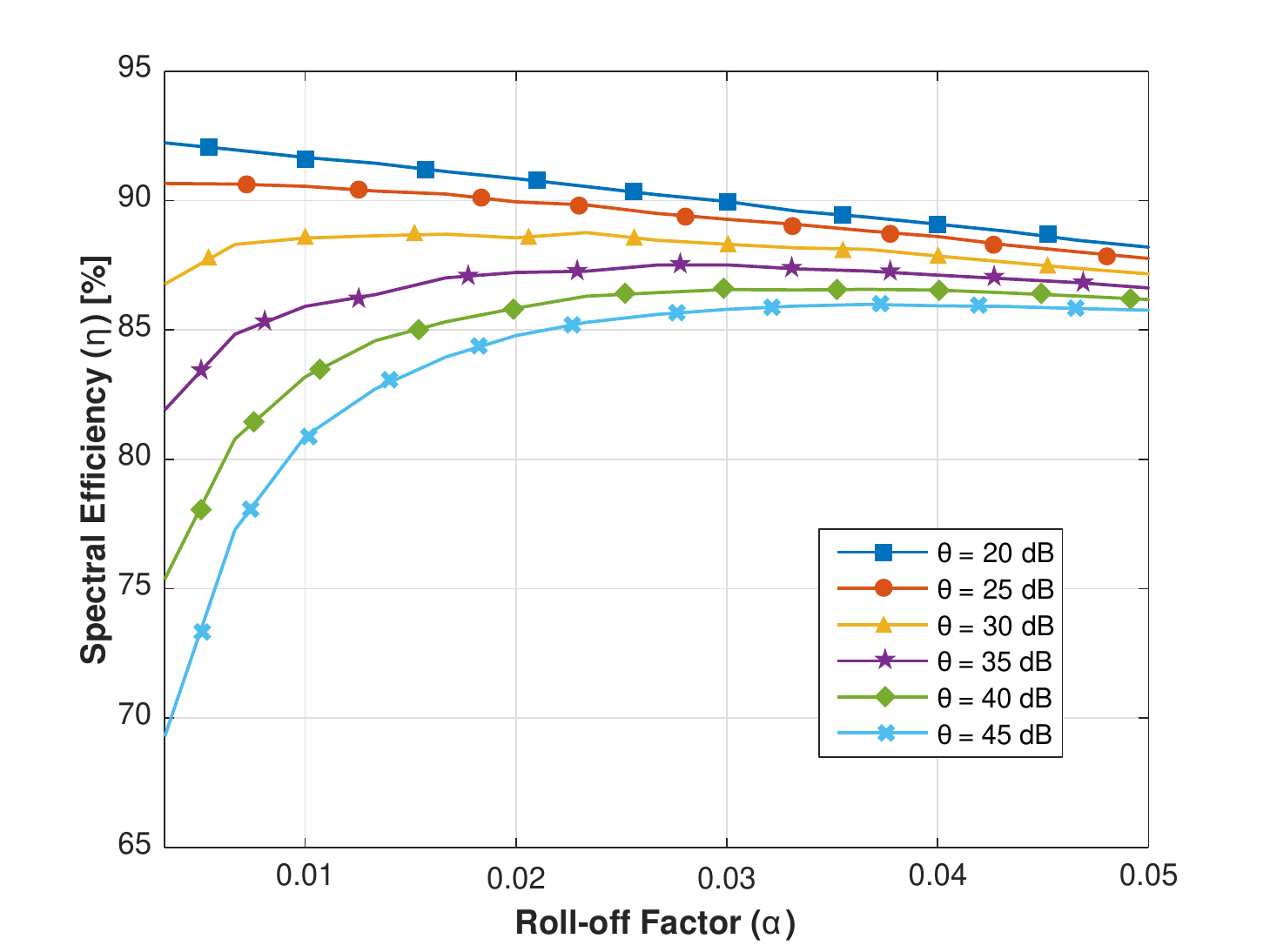}
\centering{}\caption{Spectral efficiency ($\eta$) of the GB and GD pairs that achieves
selected $\theta_{\Delta f=15\:kHz}$ (Please note that each $\alpha$
corresponds to a GB-GD pair). \label{fig:SpecEff}}
\end{figure}

\begin{table*}[b]
\caption{Optimal guard duration (GD) and guard band (GB) pairs for selected
$\theta$ \label{tab:OOBEthresh}}

\centering\includegraphics[width=1.95\columnwidth]{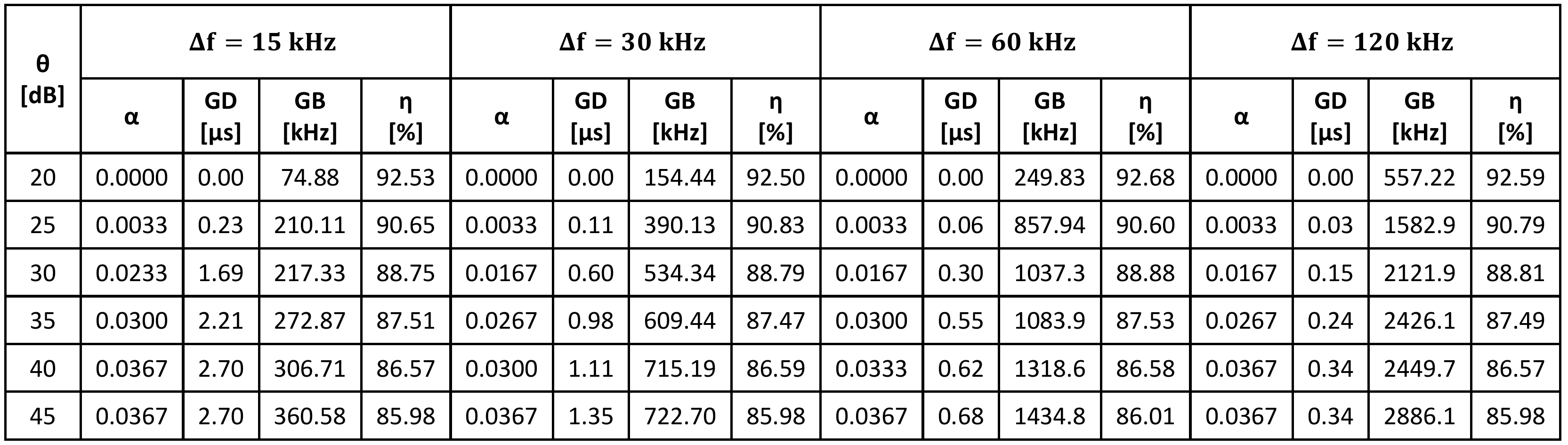}
\end{table*}

In a mixed numerology network, the INI can be managed by windowing
operation and allocating guard band between adjacent numerologies
as described in Section \ref{sec:II}. Since the windowing operation
reduces the unwanted emissions with a cost of extra guard duration,
the INI management procedure boil downs to the adaptive utilization
of guard duration (GD) and guard band (GB) to achieve a desired interference
threshold ($\theta$). Figure \ref{fig:GBvsGD} demonstrates the required
GB and GD amounts for selected $\theta$ values considering a W-OFDM
signal with $\Delta f=15\:kHz$. Each $\alpha$ value in the figure
represents a GD allocation to carry out windowing operation, and a
GB allocation to handle the rest of interference for a given $\theta$.

A tremendous time-frequency resource is required to deal with the
INI issue only with GB or GD allocation. Hence, GB and GD have to
be jointly optimized in order to improve the spectral efficiency,
which is measured as the information rate that can be transmitted
over a given bandwidth. This hyper-parameter optimization has been
carried out by a grid search method through a manually designated
subset of the hyper-parameter space \cite{bergstra2011}. The spectral
efficiency ($\eta$) is proportional to the multiplication of efficiencies
in the time and frequency domains, which are calculated as follows:

\begin{equation}
\eta_{time}=\frac{T_{OFDM}}{T_{OFDM}+T_{CP-Ch}+T_{CP-Win}}\label{eq:SE_in_Time}
\end{equation}

\begin{equation}
\eta_{freq}=\frac{OBW}{OBW+(GB\times2)}\label{eq:SE_in_Freq}
\end{equation}
Considering $T_{OFDM}$, $T_{CP-Ch}$, and occupied bandwidth ($OBW$)
are fixed parameters for a given $\Delta f$, the degrees of freedom
that can be selected independently becomes only GB and GD (i.e., $T_{CP-Win}$).
The optimization problem that looks for the optimal GB and GD pair
can be defined as follows: 

\begin{equation}
(GB,\:GD)\:=\argmax_{GB,\:GD}(\eta_{time}\times\eta_{freq})\:,\label{eq:Opt_Objective}
\end{equation}

\vspace{-5pt}

\begin{equation}
\begin{array}{cc}
\textrm{subject\:to:} & P_{i}-P_{j}+S_{j}\leq\theta_{\Delta f,i}\:.\end{array}\label{eq:Opt_Constraint}
\end{equation}

The spectral efficiencies for selected $\theta$ values are shown
in Fig. \ref{fig:SpecEff}. Each $\alpha$ value in the figure is
equivalent to a GB-GD pair for a given $\theta$, and the peak value
of each curve determines the optimal pair. These optimal pairs are
summarized in Table \ref{tab:OOBEthresh} along with the related parameters
for various $\Delta f$. The results reveal that the need for windowing
diminishes as $\theta$ decreases, and accordingly, the desired interference
level can be accomplished only with a few guard subcarriers. Also,
the spectral efficiency increases with the decrease in $\theta$.
The change in required guards clearly confirms that the adaptive guard
design enhances the spectral efficiency significantly compared to
designing the mixed numerology system considering the worst case scenario
(e.g., $\eta_{\theta=45\:dB}$ = $85.98$ \% whereas $\eta_{\theta=20\:dB}$
= $92.53$ \%). Despite the fact that the computational complexity
increases compared to traditional OFDM-based systems, the computation
of the optimal GB-GD pairs is an offline action that needs a one-time
calculation. Therefore, a lookup table procedure can be used to decrease
complexity.

\section{Inter-numerology Interference (INI)-based Scheduling\label{sec:IV}}

The total guard amount is reduced with the joint optimization of the
guard band (GB) and guard duration (GD) for a given interference threshold
($\theta_{\Delta f}$) in Section \ref{sec:III}. The optimization
results show that the spectral efficiency ($\eta$) decreases as $\theta$
increases. Also, the numerologies with larger subcarrier spacing ($\Delta f$)
require more guards, and they lead to lower $\eta$ values in a mixed
numerology network. Since $\theta$ depends on the numerologies operating
in the adjacent bands, the potential of adaptive guards can be enhanced
further along with the utilization of an interference-based scheduling
algorithm. 

\begin{figure}[!b]
\includegraphics[width=1\columnwidth]{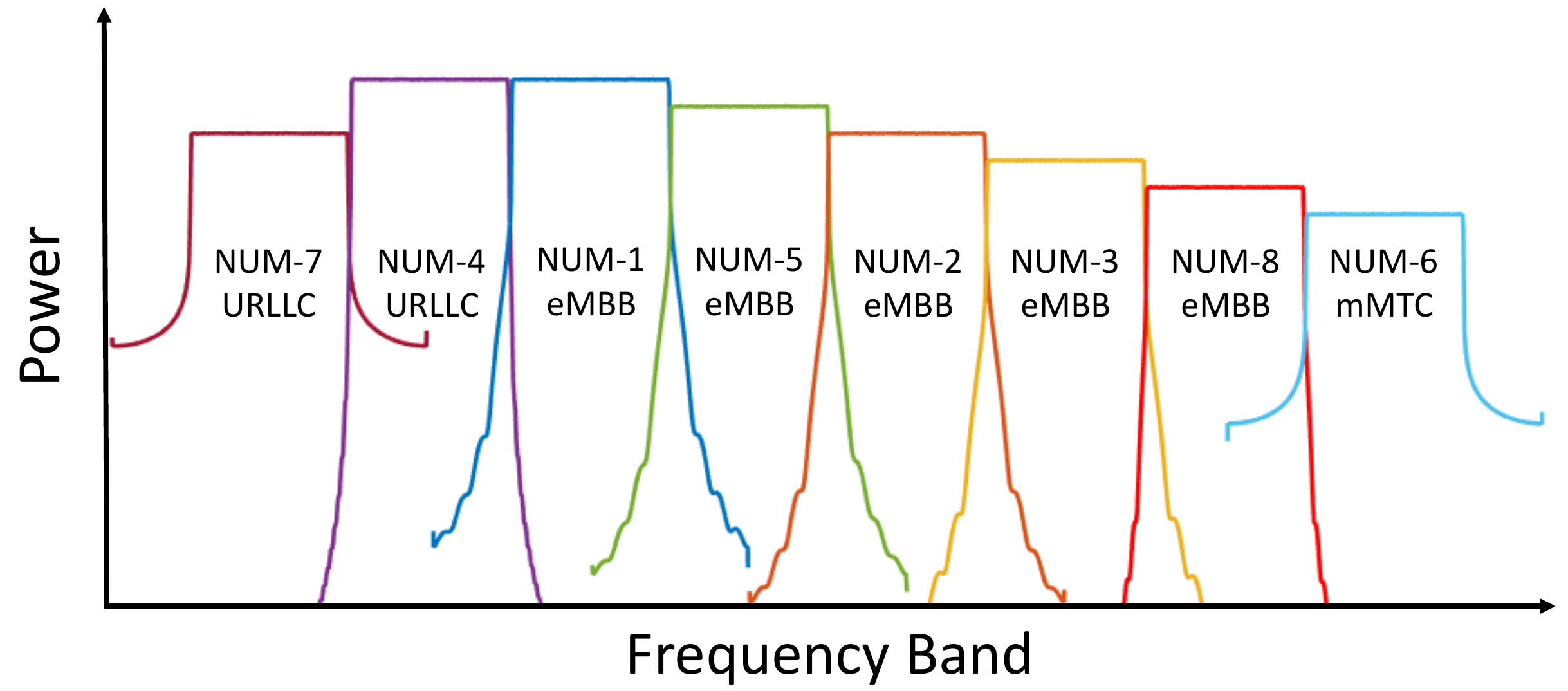}

\caption{INI-based scheduled numerologies with various use cases.\label{fig:ScheduledBands}}
\end{figure}

The proposed scheduling algorithm groups the numerologies and allocate
them to the available subbands considering the inter-numerology interference
(INI). The numerologies with similar subcarrier spacing, power level,
and SIR requirements are arranged together in order to decrease the
mean $\theta$ in the network. Consequently, the need for guards reduces,
and the spectral efficiency improves. The steps of the proposed INI-based
scheduling algorithm are listed as follows:
\begin{enumerate}
\item Sort the numerologies regarding their $\Delta f$ value in an ascending/descending
order. 
\item Calculate the similarity metric for all numerologies as $\beta_{j}=SIR_{j}-P_{j}$.
\item Sort $\beta$ in an ascending/descending order for the numerologies
with the same $\Delta f$.
\item If $\beta$ value repeats, sort based on power in the adjacent band.
\item Check $P$ on both sides of the available band. If $P$ is the same
with the numerology in its adjacent band, allocate the numerology
with a higher SIR requirement to the edge.
\end{enumerate}
The performance of the INI-based scheduling algorithm is evaluated
numerically, and its performance is compared with the performance
of a random scheduling strategy. The allocation probability to any
given subband is $1/M$ for all numerologies (i.e., uniform probability
distribution) in the random scheduling strategy. Eight numerologies
(i.e., $M=8$) are considered in Monte Carlo simulations, and a random
parameter set is assigned to numerologies for each realization, as
discussed in Section \ref{sec:II}. The key parameters of an exemplary
realization, such as subcarrier spacing, power level, and SIR requirement,
are listed in Tables \ref{tab:RandomBands} and \ref{tab:ScheduledBands}.
The random scheduling strategy and the INI-based scheduling strategy
are implemented for the same parameter set, as illustrated in Fig.
\ref{fig:Multi-window} and Fig. \ref{fig:ScheduledBands}, respectively.
Although the guards in both time and frequency domains are jointly
optimized in the PHY layer for both cases, they utilize different
scheduling algorithms in the MAC layer. As a result, any performance
difference can be attributed to the proposed scheduling algorithm.
Furthermore, a fixed guard allocation strategy is implemented with
the random scheduling method to demonstrate the effectiveness of the
proposed adaptive guard allocation. The guards are assigned considering
the worst-case scenario (i.e.,~highest~$\theta_{\Delta f}$) in
the fixed guard allocation strategy.

\begin{table}[!t]
\caption{Key parameters of randomly scheduled numerologies~for~adaptive guard
allocation\label{tab:RandomBands}}

\centering\includegraphics[width=1\columnwidth]{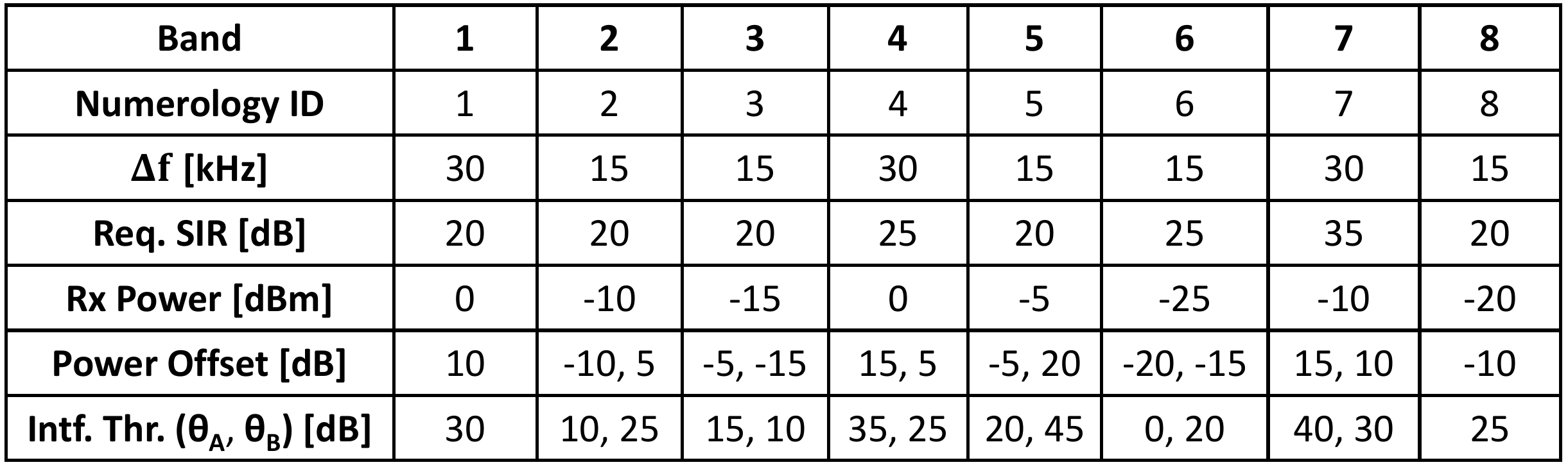}
\end{table}

\begin{table}[!t]
\caption{Key parameters of INI-based scheduled numerologies~for~adaptive
guard allocation\label{tab:ScheduledBands}}

\centering\includegraphics[width=1\columnwidth]{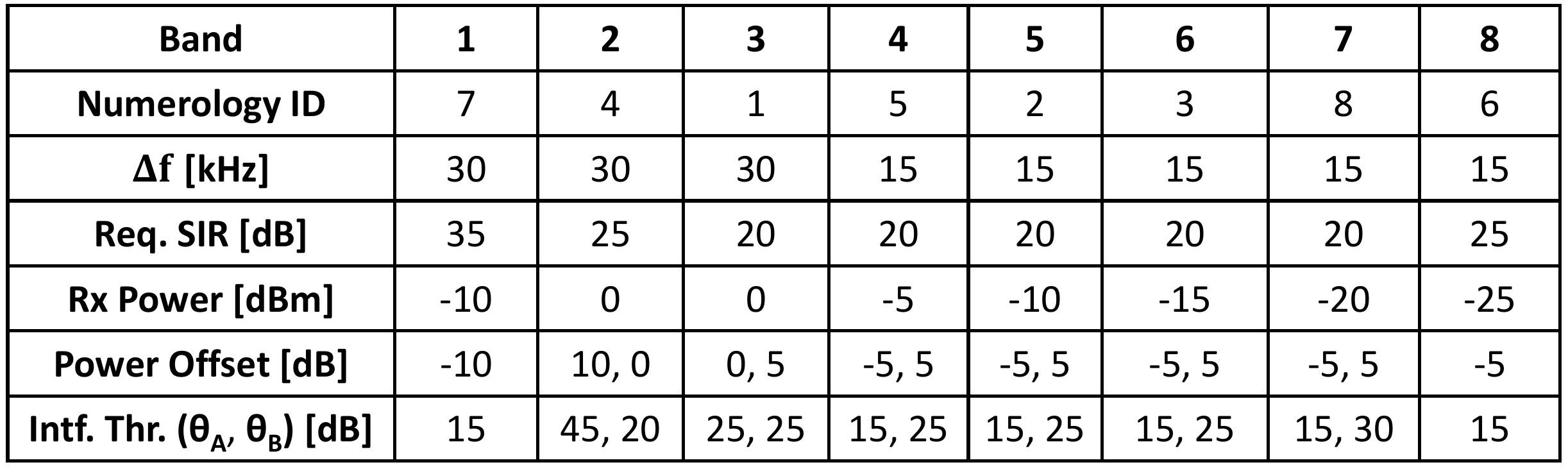}
\end{table}

The system performance is evaluated in terms of spectral efficiency
($\eta$) and the average evaluation results (out of $100$ independent
realizations) for various guard allocation and scheduling strategies,
which are the fixed guard allocation with the random scheduling, the
adaptive guard allocation with the random scheduling, and the adaptive
guard allocation with the INI-based scheduling, are presented in Table
\ref{tab:Comp}. The results demonstrate that the GD and GB amounts
are reduced by $43$\% and $34$\%, respectively when the fixed guards
are replaced with the adaptive guards in the frequency range-1 (FR1)
scenario. Also, the GD and GB amounts are reduced further by $10$\%
and $27$\%, respectively, when the proposed INI-based scheduling
strategy is implemented instead of the random scheduling strategy.
It is worth to note that $\eta$ is lower in the frequency range-2
(FR2) case since more guards are required for the numerologies with
higher $\Delta f$ values due to their higher unwanted emissions.
Although it can be compensated with an increased number of subcarriers
(FR2 is suitable for wider bands), it is kept as fixed to $256$ for
a fair comparison with the FR1 case in the numerical evaluations.

\begin{table}[!t]
\caption{Spectral efficiency comparison for various guard~allocation~and
scheduling strategies \label{tab:Comp}}

\centering\includegraphics[width=0.85\columnwidth]{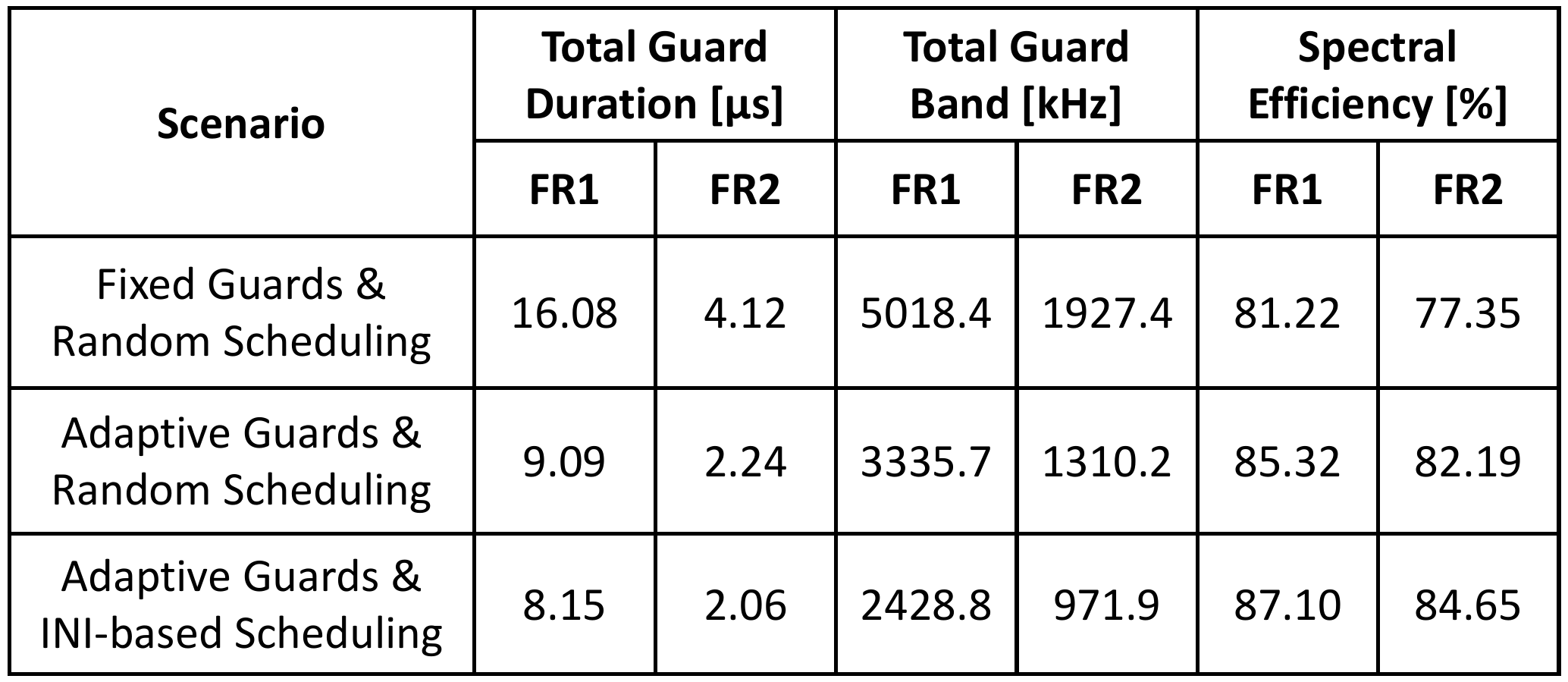}
\end{table}

The proposed INI-based scheduling strategy is particularly important
when there is a severe power imbalance between the numerologies (i.e.,
a set of synchronous UEs). The current mobile networks adopted a power
control mechanism in the uplink to manage interference between neighboring
bands. However, this solution restricts the UEs with better channel
conditions to deploy higher-order modulations and decreases the spectral
efficiency. The proposed scheduling technique can relax the power
control mechanism and improve throughput. Also, it should be noted
that the channel-based scheduling can be performed to orthogonal/synchronous
UEs within a given numerology, whereas the INI-based scheduling can
be performed to non-orthogonal/asynchronous numerologies for reduced
complexity.

\section{Conclusions\label{sec:V}}

A novel inter-numerology interference (INI) management technique with
a cross-layer approach is proposed in this paper. The adaptive guards
in both time and frequency domains are utilized along with a multi-window
operation in the PHY layer and jointly optimized considering the use
case, subcarrier spacing, and power offset between the numerologies.
Since the allowed interference level depends on the numerologies operating
in the adjacent bands, the potential of adaptive guards is further
increased and exploited with a MAC layer scheduling technique. The
proposed INI-based scheduling algorithm decreases the need for guards
by allocating the numerologies to the available bands, considering
their subcarrier spacing, power level, and SIR requirements. It is
demonstrated that the optimized guard allocation and INI-based scheduling
algorithm improves the spectral efficiency significantly while taking
into account the different use case requirements and device capabilities. 

The results show that the precise design that accommodates such flexibility
reduces the guards and improves the performance of mixed numerology
systems. The INI management technique is proposed on the transmitter
side in the PHY layer along with a MAC layer scheduling technique
in this study. The guards are designed in such a way that it guarantees
the required SIR levels for each numerology in the network. Therefore,
the theoretical upper bounds on the bit error rate can be obtained
in a straightforward way using the channel capacity equation \cite{proakis2001}.
However, it will be extended to the receiver side as well for enhanced
performance in the future. Also, a practical receiver structure enables
performance evaluation under various channel conditions and impairments.
Furthermore, the CP length for the multipath channel is assumed fixed
and sufficient for the delay spread. Nevertheless, some ISI can be
allowed in order to suppress INI further for a fixed guard duration,
and the ISI vs. INI trade-off is worth investigating. Last but not
least, the proposed INI management technique with a cross-layer approach
can be applied to other spectrally enhanced OFDM systems \cite{zhang2015,vakilian2013,fettweis2009,farhang2016}
as well by optimizing the waveform parameters and guards along with
a proper scheduling mechanism. For example, a recent publication \cite{yang2019}
demonstrated the joint optimization of the filter parameters and guard
band for filtered-OFDM (f-OFDM) only in the PHY layer. 

The next-generation wireless communication technologies are evolving
towards increased flexibility in various aspects. Enhanced flexibility
is the key design consideration, especially to be able to serve diverse
requirements. Hence, the adaptive guard utilization must be a part
of the future communication systems.

\bibliographystyle{IEEEtran}
\bibliography{IEEEabrv,WaveformRef}

\vspace{-5cm}
\begin{IEEEbiography}[\includegraphics{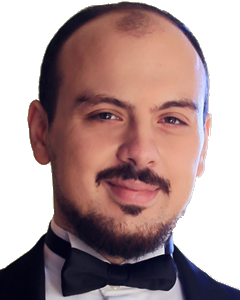}]{Ali Fatih Demir}
(S'08) received the B.S. degree in electrical engineering from Y\i ld\i z
Technical University, Istanbul, Turkey, in 2011 and the M.S. degrees
in electrical engineering and applied statistics from Syracuse University,
Syracuse, NY, USA in 2013. He is currently pursuing the Ph.D. degree
as a member of the Wireless Communication and Signal Processing (WCSP)
Group in the Department of Electrical Engineering, University of South
Florida, Tampa, FL, USA. His current research interests include PHY
and MAC aspects of wireless communication systems, wireless body area
networks (WBANs), and signal processing/machine learning algorithms
for brain-computer interfaces (BCIs).
\end{IEEEbiography}

\vspace{-4.5cm}
\begin{IEEEbiography}[\includegraphics{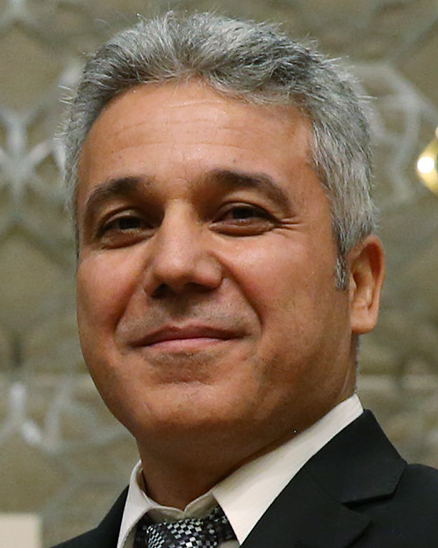}]{H{\"u}seyin Arslan}
 (S\textquoteright 95\textendash M\textquoteright 98\textendash SM\textquoteright 04\textendash F\textquoteright 16)
received the B.S. degree in electrical and electronics engineering
from Middle East Technical University, Ankara, Turkey, in 1992, and
the M.S. and Ph.D. degrees in electrical engineering from Southern
Methodist University, Dallas, TX, USA, in 1994 and 1998, respectively.
From January 1998 to August 2002, he was with the research group of
Ericsson Inc., Charlotte, NC, USA, where he was involved with several
projects related to 2G and 3G wireless communication systems. Since
August 2002, he has been with the Department of Electrical Engineering,
University of South Florida, Tampa, FL, USA, where he is a Professor.
In December 2013, he joined Istanbul Medipol University, Istanbul,
Turkey, where he has worked as the Dean of the School of Engineering
and Natural Sciences. His current research interests include waveform
design for 5G and beyond, physical layer security, dynamic spectrum
access, cognitive radio, coexistence issues on heterogeneous networks,
aeronautical (high altitude platform) communications, and \emph{in
vivo} channel modeling and system design. He is currently a member
of the editorial board for the \emph{Sensors Journal} and the \emph{IEEE
Surveys and Tutorials}.
\end{IEEEbiography}

\end{document}